# A microstructurally informed model for the mechanical response of three-dimensional actin networks

Ronald Y. Kwon[*,**,##], Adrian J. Lew[##], Christopher R. Jacobs[**,##]

[##] Department of Mechanical Engineering, Stanford University, Stanford, CA, 94305
[**] Bone and Joint Rehabilitation R&D Center, Department of Veterans Affairs, Palo Alto, CA, 94304
[*] Corresponding author. Email: ronkwon@stanford.edu

We propose a class of microstructurally informed models for the linear elastic mechanical behavior of cross-linked polymer networks such as the actin cytoskeleton. Salient features of the models include the possibility to represent anisotropic mechanical behavior resulting from anisotropic filament distributions, and a power-law scaling of the mechanical properties with the filament density. Mechanical models within the class are parameterized by seven different constants. We demonstrate a procedure for determining these constants using finite element models of three-dimensional actin networks. Actin filaments and cross-links were modeled as elastic rods, and the networks were constructed at physiological volume fractions and at the scale of an image voxel. We show the performance of the model in estimating the mechanical behavior of the networks over a wide range of filament densities and degrees of anisotropy.

*Keywords:* Homogenization; Constitutive model; Finite element model; Actin cytoskeleton; Cell mechanics

*AMS Subject Classification:* 74D05; 74K10; 74S05; 74E10

## 1 Introduction

Numerous experiments have shown mechanical loading to be an important factor in the development and/or maintenance of a wide variety of tissues such as muscle, cartilage, tendon, and bone, and organs such as the heart and lung. The deformations of the cells within these tissues and organs are dictated by their mechanical behavior under loading. Thus, it comes as no surprise that cellular mechanical behavior has been implicated as an important factor in the pathology of many diseases such as osteoporosis, osteoarthritis, cancer, heart failure, and several pulmonary disorders [1].

Our understanding of the mechanical regulation of the pathologic processes involved in these diseases would be greatly enhanced if it were possible to predict the mechanical behavior of a particular cell from microscopically obtained observations. A critical component governing the mechanical behavior of adherent cells is the actin cytoskeleton, a three-dimensional network of cross-linked actin filaments (figure 1). The microstructure of the actin cytoskeleton is highly dynamic and can change dramatically in response to mechanical loading. A growing body of evidence suggests that the ability of cells to convert mechanical signals into biochemical signals depends on actin cytoskeletal microstructure [2, 3]. Microstructurally-based models of the actin cytoskeleton would be ideal for investigating the mechanical implications of actin microstructural organisation, since representative cytoskeletal networks observed *in vitro* could be examined. Important microstructural features, such as spatial and angular heterogeneity, could be directly accounted for, allowing investigation of underlying mechanical 'principles' that may be governing cytoskeletal microarchitecture.

Homogenized models for the actin cytoskeleton have yet to be obtained. A homogenized model in this context is a constitutive model for a continuum capable, in some appropriate sense, of approximating the mechanical behavior of the network. The advantage of a homogenized model is that a complete description of the network microstructure is not needed to specify its mechanical behavior. Although some advanced results are available for structured networks [4], rigorous mathematical results on general network homogenization problems remain elusive. Recent and classical theoretical investigations of 'stiff' or 'semi-flexible' polymer networks have yielded important insight into the mechanics of this class of networks and have generally identified geometric properties of random networks [5, 6, 7], different elastic regimes [8], scaling behaviors, and methods for explicit calculation of the macroscopic network elastic moduli from microscopic properties [9, 10, 11, 12, 13, 14, 15, 16, 17, 18, 19, 20, 21, 22, 23, 24, 25, 26, 27, 28, 29, 30]. All of these results but [24] have been obtained for two-dimensional, isotropic networks. However, anisotropy is highly relevant for actin networks, which form aligned bundles of actin within the cytoskeleton (both in cultured cells and *in vivo* [31]) as well as cross-linked gels [32, 33]. The direct consequence of anisotropy is that, as opposed to previous investigations that needed to solely focus on the tensile and/or shear modulus that characterize isotropic materials, a suitable homogenized model will require the calculation of the entire elastic tensor (21 independent components) to fully specify the mechanical behavior of the network.

Incomplete knowledge of microscopic network properties is a unique challenge which makes constitutive modeling of many biopolymer networks difficult. In the case of the actin cytoskeleton, although the dimensions and material properties of individual actin filaments have been measured [34], many important microscopic properties of actin cytoskeletal networks have yet to be elucidated. For example, there are a wide variety of cytoskeletal cross-linking proteins *in vivo* whose mechanical behavior need to be characterized. In addition, although it is well accepted that the cytoskeletal network can be subject to a prestress, the degree to which each filament is prestressed is not known.

We propose here a novel class of models for the homogenized linear elastic response of cross-linked polymer networks such as the actin cytoskeleton. The proposed models can be constructed based on the filament angular distribution and spatial density. We deliberately avoided making specific assumptions on whether the elasticity of the network is the result of entropic [11] or enthalpic [8, 15] contributions, the nature of the cross-links between filaments in the network, or the existence of a prestress in the network, since these are still largely under discussion. Instead, our goal was to formulate a class of models that account for some features of the microstructure of the network, and that, through a suitable validation procedure, could be tailored to represent its homogenized linear elastic response under any or many of these conditions.

The 21 elastic moduli of the model are determined by postulating an ansatz or functional form inspired from the exact expression for affinely deformed networks with anisotropic filament distributions, see e.g., [35, 36]. The model accounts for the possibly different-than-linear exponents observed in the power law dependence of the elastic moduli with filament density (see, e.g., [8, 15]), and the effect of cross-links in the Poisson ratio. There are only 7 independent parameters, which need to be calibrated from a relatively small number of simulations of fully resolved and explicitly represented networks. We showcase the performance of the model by predicting the mechanical response of finite element models of three-dimensional, anisotropic networks of elastic rods with semi-flexible cross-links. The calibrated model shows good performance over a wide range of angular distributions and spatial densities away from the vicinity of the point of calibration. The particular type of networks chosen for this example was motivated by two-dimensional analogs that have been previously adopted as possible descriptions for the actin cytoskeleton [13, 15]. We expect, however, that the homogenized class of models proposed herein will also be useful to express the effective behavior of more general network types, resulting from a future enhanced understanding of key features of the actin cytoskeleton.

Throughout, vectors are denoted by boldface lowercase latin characters, second-order tensors by boldface lowercase greek characters, and fourth-order tensors by boldface uppercase latin characters. All tensor components are referred to an orthonormal basis. All nonboldface characters are considered scalar quantities. When indicial notation is used, an index appearing twice in a term indicates sum over it in the range 1 to 3.

## 2 Model Formulation

We propose and detail here a class of models to approximate the homogenized linear elastic response of cross-linked polymer networks such as the actin cytoskeleton. We idealize network filaments as cylindrical rods of cross-sectional area $A$ with extensional, bending, and torsional stiffness. The configuration of each filament can be characterized by its midpoint position and direction in space when unstressed. We consider three-dimensional, infinite networks that are periodic with period $L$ in three orthogonal directions. The unit cell of the network is then a cubic box of side $L$. A cross-link between two filaments may be formed whenever the distance between the two is within a specified distance. We denote with $\rho$ the volume fraction of filaments in the network, i.e., the quotient between the total volume occupied by all filaments and the volume of the unit cell $L^3$. Additionally, let $\omega(\boldsymbol{n})$ be the angular probability density of the volume fraction, which indicates the angular distribution of the volume fraction $\rho$. The value of $\omega(\boldsymbol{n})\rho dS$ gives the volume fraction of filaments with orientation in a neighborhood $dS$ of the unit sphere surrounding $\boldsymbol{n}$. It satisfies that its integral over the unit sphere $S^2$ is equal to one.

Motivated by homogenization results for linear periodic composite materials (see e.g., [37]), we expect the elastic energy density of the homogenized model to be well approximated by the lowest elastic energy attainable by the network under suitably imposed boundary conditions on the unit cell. For this study, the points of intersection between the network filaments and the unit cell faces are constrained to follow an affine deformation. A class of less stringent boundary conditions requiring only periodic but not necessarily affine displacements on the boundary of the unit cell are also of interest, but will not be considered here. If $\boldsymbol{x}$ denotes the position vector of one such point and $\boldsymbol{\varepsilon}$ is a second-order tensor, then an affine deformation $\mathcal{T}_{\boldsymbol{\varepsilon}}(\boldsymbol{x})$ maps $\boldsymbol{x}$ to $\boldsymbol{x}+\boldsymbol{\varepsilon}\cdot\boldsymbol{x}$. Since we are first interested in the linearized elastic behavior of initially stress-free networks, only affine deformations in which $\boldsymbol{\varepsilon}$ is small and symmetric need to be considered. In this case, the elastic energy density of the homogenized models has the form $w(\boldsymbol{\varepsilon})=C_{ijkl}\varepsilon_{ij}\varepsilon_{kl}/2$, where $C_{ijkl}$ are the Cartesian components of the homogenized or effective elastic moduli. The elastic moduli should satisfy that $C_{ijkl}=C_{klij}=C_{jilk}$ (see, e.g., [38]), also known as major and minor symmetries, respectively.

Even though homogenization of the mechanical properties for general networks is still an open problem, there are some cases in which the elastic moduli are easily obtained. Consider for example the case of an initially unstressed network in which every filament has only extensional stiffness $K_{//}$ (no bending or torsion). All cross-links in the network are assumed to be able to freely rotate with no energetic cost. It is easily seen that upon constraining the filament ends to follow an affine deformation $\mathcal{T}_{\boldsymbol{\varepsilon}}$, an equilibrium configuration of the network is obtained by mapping all cross-links with $\mathcal{T}_{\boldsymbol{\varepsilon}}$ as well (in the linear elastic regime of concern here, such network deformation has the lowest energy among all those that use the same boundary conditions; it may not be unique though). Herein, we call this class of networks 'affine networks'. As we shall detail later, the homogenized elastic moduli in this case are given by

$$C_{ijkl} = K_{//} \int_{S^2} \rho\omega(\boldsymbol{n})n_i n_j n_k n_l dS \; . \qquad (1)$$

As nicely discussed in [12], this result will not strictly hold when the filament ends are not constrained to follow an affine deformation. Previous numerical studies have also shown that the linear scaling of the moduli with $\rho$ is not valid for more complex networks, such as when bending stiffness is accounted for. Instead, power laws of the form $(\rho-\rho_{ref})^\alpha$ for some exponent $\alpha\geq1$ and reference density $\rho_{ref}$ have been found to give a good approximation to the dependence of the shear or Young modulus with the density for isotropic networks, over spans of several orders of magnitude [10, 15]. The value of $\rho_{ref}$ is typically related to the percolation threshold, e.g., [23, 24].

We are therefore motivated to postulate the following form for the homogenized elastic moduli

$$C_{ijkl} = \int_{S^2} (\rho - \rho_{ref})^\alpha \, \omega(\boldsymbol{n})K_{ijkl}(\boldsymbol{n})dS \; . \qquad (2)$$

Here, $K_{ijkl}(\boldsymbol{n})$ is a fourth-order tensor valued function of the direction vector $\boldsymbol{n}$, while $\rho_{ref}$ and $\alpha$ are scalars, all of them to be determined, and we consider only volume fractions $\rho > \rho_{ref}$. In the particular case of affine networks in equation (1), we have $\rho_{ref}=0$, $\alpha=1$ and

$$K_{ijkl}(\boldsymbol{n}) = K_{//}n_i n_j n_k n_{l.} \tag{3}$$

The homogenized elastic moduli in equation (2) should have major and minor symmetries for *any* choice of angular density $\omega(\boldsymbol{n})$. It then follows that $\boldsymbol{K}(\boldsymbol{n})$ should also obey these same symmetries, and hence $K_{ijkl}=K_{klij}=K_{jilk}$. This leaves at most 21 independent components of $\boldsymbol{K}(\boldsymbol{n})$ to be determined, all of them functions over the unit sphere. However, we show next that these components cannot be arbitrary functions of $\boldsymbol{n}$, but that they are fully determined by five independent constants and a specific functional dependence on $\boldsymbol{n}$. The elegant derivation below significantly reduces the complexity of the model and sheds light into the physical interpretation of $\boldsymbol{K(n)}$.

A physically reasonable requirement on the elastic moduli $C_{ijkl}$ is that if the entire network is rigidly rotated by a rotation $\boldsymbol{Q}$ then the elastic moduli should rotate accordingly, also known as the material frame indifference requirement in continuum mechanics, see e.g., [39]. More precisely, after rotating the network by $\boldsymbol{Q}$, a new network is obtained with an angular probability density $\underline{\omega}(\boldsymbol{n})=\omega(\boldsymbol{Q}^{-1}\boldsymbol{n})$ and elastic moduli

$$\begin{aligned}
\underline{C}_{ijkl} &= \int_{S^2} (\rho - \rho_{ref})^\alpha \, \underline{\omega}(\boldsymbol{n}) K_{ijkl}(\boldsymbol{n}) dS \\
&= \int_{S^2} (\rho - \rho_{ref})^\alpha \, \omega(\boldsymbol{Q}^{-1}\boldsymbol{n}) K_{ijkl}(\boldsymbol{n}) dS \\
&= \int_{S^2} (\rho - \rho_{ref})^\alpha \, \omega(\boldsymbol{n}) K_{ijkl}(\boldsymbol{Qn}) dS \, .
\end{aligned} \tag{4}$$

Correspondingly, by material frame indifference the elastic moduli of the rotated network should be given by $\underline{C}_{IJKL}=Q_{Ii}Q_{Jj}Q_{Kk}Q_{Ll}C_{ijkl}$, or equivalently,

$$\underline{C}_{IJKL} = Q_{Ii}Q_{Jj}Q_{Kk}Q_{Ll} \int_{S^2} (\rho - \rho_{ref})^\alpha \, \omega(\boldsymbol{n}) K_{ijkl}(\boldsymbol{n}) dS \, . \tag{5}$$

Equating Eqs. 4 and 5, and utilizing the fact that the resulting identity should be valid for any filament angular probability density $\omega(\boldsymbol{n})$, we obtain

$$K_{IJKL}(\boldsymbol{Qn}) = Q_{Ii}Q_{Jj}Q_{Kk}Q_{Ll}K_{ijkl}(\boldsymbol{n}) \, , \tag{6}$$

which precisely defines the dependence of $\boldsymbol{K}$ on $\boldsymbol{n}$. Notice that determining $\boldsymbol{K}$ in a single direction $\boldsymbol{n}_0$ allows its calculation in any other direction $\boldsymbol{n}$ through an appropriately chosen rotation $\boldsymbol{Q}$ that maps $\boldsymbol{n}_0$ to $\boldsymbol{n}$. In fact, there are an infinite number of such rotations. If $\boldsymbol{Q}$ denotes one such rotation, then any of the others is obtained as $\boldsymbol{QR}$, where $\boldsymbol{R}$ is any rotation about $\boldsymbol{n}_0$. This implies that for any such $\boldsymbol{R}$, with $\boldsymbol{Q}$ equal to the identity,

$$K_{IJKL}(\boldsymbol{n}_0) = R_{Ii}R_{Jj}R_{Kk}R_{Ll}K_{ijkl}(\boldsymbol{n}_0) \; , \qquad (7)$$

or alternatively, the tensor $\boldsymbol{K}(\boldsymbol{n}_0)$ must be transversely isotropic about $\boldsymbol{n}_0$ (see e.g., [40]). There are consequently only five independent parameters in $\boldsymbol{K}(\boldsymbol{n}_0)$. These values are easily expressed in the Voigt basis [40], the 6x6 matrix representation of $\boldsymbol{K}(\boldsymbol{n}_0)$. Assuming that the axis of symmetry $\boldsymbol{n}_0$ is parallel to $\boldsymbol{e}_3$, the unit vector parallel to the third coordinate axis, we get

$$\boldsymbol{K}(\boldsymbol{e}_3){=}\begin{bmatrix} K_{1111}(\boldsymbol{e}_3) & K_{1122}(\boldsymbol{e}_3) & K_{1133}(\boldsymbol{e}_3) & 0 & 0 & 0 \\ K_{1122}(\boldsymbol{e}_3) & K_{1111}(\boldsymbol{e}_3) & K_{1133}(\boldsymbol{e}_3) & 0 & 0 & 0 \\ K_{1133}(\boldsymbol{e}_3) & K_{1133}(\boldsymbol{e}_3) & K_{3333}(\boldsymbol{e}_3) & 0 & 0 & 0 \\ 0 & 0 & 0 & \frac{1}{2}(K_{1111}(\boldsymbol{e}_3)-K_{1122}(\boldsymbol{e}_3)) & 0 & 0 \\ 0 & 0 & 0 & 0 & K_{1313}(\boldsymbol{e}_3) & 0 \\ 0 & 0 & 0 & 0 & 0 & K_{1313}(\boldsymbol{e}_3) \end{bmatrix}. \; (8)$$

Here $K_{1111}(\boldsymbol{e}_3)$, $K_{1122}(\boldsymbol{e}_3)$, $K_{1133}(\boldsymbol{e}_3)$, $K_{3333}(\boldsymbol{e}_3)$, and $K_{1313}(\boldsymbol{e}_3)$ are the five independent constants to be determined. Once these constants are known, the entire tensor valued function over the unit sphere $\boldsymbol{K}(\boldsymbol{n})$ can be calculated using equation (6). Together with $\rho_{ref}$ and $\alpha$, this makes a total of seven constants in equation (2) needed to completely specify the model.

In order to gain insight into the physical origin of the five constants which make up $\boldsymbol{K}(\boldsymbol{n})$, we revisit the case of affine networks and compute the homogenized moduli in equation (1) and $\boldsymbol{K}$ in equation (3) for this class of networks next. The elastic energy of a filament per unit filament volume is

$$W_{filament}(\lambda) = \frac{1}{2}K_{//}\lambda^2 \; , \qquad (9)$$

where $\lambda$ is the axial strain. In the linear elastic regime and under an affine deformation $\mathscr{T}_{\boldsymbol{\varepsilon}}$ the value of $\lambda$ for a filament oriented in a direction $\boldsymbol{n}$ can be computed as $n_r\varepsilon_{rs}n_s$. The elastic energy density of the homogenized model is the result of adding the elastic energies contributed by each filament in the network divided by the volume of the unit cell, i.e.,

$$w(\boldsymbol{\varepsilon}) = \int\limits_{S^2} \rho\omega(\boldsymbol{n})W_{filament}(n_i\varepsilon_{ij}n_j)dS \; . \qquad (10)$$

The homogenized elastic moduli then follow as

$$C_{ijkl} = \frac{\partial^2 w(\boldsymbol{\varepsilon})}{\partial\varepsilon_{ij}\partial\varepsilon_{kl}} = K_{//}\int\limits_{S^2}\rho\omega(\boldsymbol{n})n_in_jn_kn_l dS \; , \qquad (11)$$

which shows the origin of equation (1) and subsequently equation (3). Upon inspection of equation (3), we see that for the affine case, $K_{3333}(\boldsymbol{e}_3){=}K_{//}$, which explicitly shows the connection between $K_{3333}(\boldsymbol{e}_3)$ and the extensional stiffness of the filaments. In fact, $K_{3333}(\boldsymbol{e}_3)$ is the only one out of the five different constants in $\boldsymbol{K}(\boldsymbol{e}_3)$ that is different from zero. Recall that this result is a direct consequence of the fact that the cross-links are mapped affinely. Thus, the nonzero values of the other coefficients could serve as indicators of the degree to which the cross-links are mapped affinely when $\boldsymbol{K}(\boldsymbol{e}_3)$ is computed for more general networks of filaments with extensional stiffness only.

These last observations provide the basis for an interesting remark about the model. The material frame indifference argument showed that the proposed ansatz is obtained by arranging a transversely isotropic material with elastic moduli $\boldsymbol{K}$ in different directions. For affine networks, the cross-links do not play a role in the deformation, i.e., they can be removed and the network deforms in the same way. This is reflected in the elastic moduli $\boldsymbol{K}$, since only the extensional stiffness in the direction of the symmetry axis, the direction of the filament, is different from zero. However, the model can admit more general moduli $\boldsymbol{K}$ when the other four constants are also different from zero. The presence of nonaffinely mapped cross-links takes advantage of this generality. In this case, in addition to the extensional stiffness provided by filaments in the direction of the axis of symmetry, $\boldsymbol{K}$ has a contribution of the bulk network. For example, by pulling in the direction of the axis of symmetry, a contraction in the transverse directions may be obtained, i.e., a nonzero Poisson ratio (figure 2). The model is then able to represent the fact that each filament transmits forces along its own direction and, through its connections to the rest of the network, in transversal directions as well. This observation suggests that the proposed model has the best chances to be accurate when each filament encounters a similar type of cross-linking with the rest of the network regardless of its orientation, for example, a similar number of cross-links. If this were not the case, the transverse mechanical properties in $\boldsymbol{K}$ would be orientation dependent, in direct contradiction of equation (6).

Simple examination reveals some of the shortcomings of the proposed model. For example, consider an initially unstressed network formed by filaments whose elastic energy arises solely due to bending (no torsion or stretching) and whose cross-links are rigid, i.e., the angle between any two filaments at a cross-link cannot change. The filaments are arranged in a prismatic microstructure so that filament directions are parallel to three orthonormal vectors $\boldsymbol{e}_1$, $\boldsymbol{e}_2$, and $\boldsymbol{e}_3$, while the distance between consecutive cross-links in each direction, are $L_1$, $L_2$ and $L_3$, respectively. As we shall show next, the homogenized linear elastic moduli for this network depend on quadratic moments of $\omega(\boldsymbol{n})$ (i.e., products of the type $\omega(\boldsymbol{e}_i)\omega(\boldsymbol{e}_j)$) which account for the number of intersections between filaments in directions $\boldsymbol{e}_i$ and $\boldsymbol{e}_j$. Clearly, this quadratic dependence on $\omega(\boldsymbol{n})$ is not represented in the model, so we do not expect the ansatz to be predictive for networks in which most of the elastic energy is used to bend the filaments

For the sake of simplicity, we shall compute only the shear modulus $G=C_{1212}$. Notice then that the periodicity of the microstructure enables us to consider a minimal unit cell with dimensions $L_1$, $L_2$ and $L_3$, which substantially simplifies the computation. We are interested in the energy of the network when an affine deformation $\mathcal{T}_\varepsilon$, with $\boldsymbol{\varepsilon}=\gamma(\boldsymbol{e}_1\otimes\boldsymbol{e}_2+\boldsymbol{e}_2\otimes\boldsymbol{e}_1)/2$ for some small constant $\gamma>0$, is used to map the positions of the cross-links in the unit cell, and hence all cross-links in the network. It is straightforward to check that by allowing the filaments to bend so that the resultant torque at each cross-link is identically zero, an equilibrium configuration of the network is obtained. For the boundary conditions under consideration, only the filaments parallel to $\boldsymbol{e}_1$ and $\boldsymbol{e}_2$ bend, and they do it within the $\boldsymbol{e}_1$-$\boldsymbol{e}_2$ plane (figure 3). The bending energy per unit length for each one of these filaments is given by $K_\perp\kappa^2/2$, where $\kappa$ denotes the local curvature and $K_\perp$ the bending stiffness. The linear elastic energy per unit volume is then easily computed to be

$$w(\gamma) = \frac{1}{L_1 L_2 L_3} \frac{24\gamma^2 K_\perp}{(L_1 + L_2)}. \tag{12}$$

The volume fraction of the network is obtained as $\rho = A(L_1+L_2+L_3)/(L_1 L_2 L_3)$, while the fraction of the total filament volume occupied by filaments parallel to $\boldsymbol{e}_i$ is denoted with $\omega_i = L_i/(L_1+L_2+L_3)$. It then follows that

$$w(\gamma) = \frac{\omega_1 \omega_2 \omega_3}{\omega_1 + \omega_2} \frac{24\gamma^2 K_\perp}{A^2}, \tag{13}$$

and that

$$G = \frac{\partial^2 w(\gamma)}{\partial \gamma^2} = \rho^2 \frac{\omega_1 \omega_2 \omega_3}{\omega_1 + \omega_2} \frac{48 K_\perp}{A^2}. \tag{14}$$

The last equation clearly shows that the stiffness for this class of networks depends on the product between values of the angular probability distribution in different directions. Additionally, the dependence of $G$ on the angular probability distribution is highly nonlinear. These two facts are in direct contradiction to the proposed model in equation (2), which explicitly contains a linear dependence on $\omega(\boldsymbol{n})$. Notice, however, that the model is able to capture the nonlinear stiffening of the network with the volume fraction.

## 3 Numerical methods

Generally, $\rho_{ref}$, $\alpha$, and the five independent components of $\boldsymbol{K}(\boldsymbol{n})$ cannot be analytically solved for. We outline here a procedure to numerically determine these parameters. We demonstrate the validity of this procedure and the proposed model by performing numerical experiments with finite element models of cross-linked networks of a specific type. By using finite element models, we can solve for the homogenized elastic moduli of each network exactly, and thus have a 'gold standard' to which we can directly compare elastic moduli calculated using equation (2).

### 3.1 Finite element modeling and computation of elastic moduli

Finite element models of three-dimensional, periodic, cross-linked networks were constructed by placing filaments of length 350nm inside cubic domains of length 400nm (figure 4). These length scales were selected since they may be relevant for image-based finite element models of the actin cytoskeleton (i.e., finite element models meshed directly from three-dimensional image data). For example, a typical pixel/voxel size in fluorescent microscopy is 0.2~0.4μm, and actin filaments are approximately 0.2~0.5μm in length *in vivo* [41, 42, 43]. Filament centroids were randomly sampled from a uniform distribution in the cubic domain, while distributions with different degrees of anisotropy were used for the orientations. We idealized actin filaments as straight, three-dimensional elastic rods. Without loss of generality, we assumed the rods to have circular cross sections. Actin filaments were modeled as Euler-Bernoulli beams with linear elastic extensional and torsional stiffness, with a diameter $d$=8nm and Young modulus $E$=1.8GPa [34]. There are a large number of different cross-linking proteins *in vivo* with a wide variety of lengths and microscopic properties (many of which have yet to be elucidated). For simplicity, we assumed that actin filaments with centerlines (i.e., axes of symmetry) closer than a rod diameter to each other were cross-linked by straight elastic rods with circular cross-sections. The cross-link rod length was the distance between the centerlines of two filaments being cross-linked, and the cross-link ends were rigidly fixed to the filaments being cross-linked. The cross-links were assumed to have similar cross-sectional diameter ($d$=8nm) and material properties ($E$=1.8GPa) as the actin filaments. Note that our network construction is similar to that of three-dimensional soft-core fiber networks [44], in that the filaments were allowed to completely penetrate one another, and overlapping filaments were considered cross-linked. Of direct consequence is that unlike two-dimensional networks (e.g., [8, 15]), the rod diameter played a critical role in the network connectivity, since the probability of a fixed number of filaments to intersect asymptotically vanishes with the rod diameter [45]. Once the filaments were placed in the cubic domain, filament segments that extended beyond the domain were reintroduced back into the domain as if the cube was the unit cell of a periodic network, similar to [46]. We imposed periodic boundary conditions on the walls of the unit cell by constraining the points of intersection between filaments and the cubic domain boundary to follow a prescribed affine deformation, while the torques at the same points were constrained to be continuous across the two parts of the filament lying in neighboring unit cells. The total strain energy density for a given deformation $\varepsilon$ imposed on the boundary was computed using the finite element method (ABAQUS Standard 6.4-1; Abaqus Inc., Providence, RI; static analysis; B31 elements). Once the strain energy densities were found, the homogenized network elastic moduli were calculated as

$$C_{ijkl} = \frac{\partial^2 w}{\partial \varepsilon_{ij} \partial \varepsilon_{kl}}(\mathbf{0}) \approx \frac{[w(\Delta \mathbf{e}_i \otimes \mathbf{e}_j + \Delta \mathbf{e}_k \otimes \mathbf{e}_l) - w(\Delta \mathbf{e}_i \otimes \mathbf{e}_j)] - [w(\Delta \mathbf{e}_k \otimes \mathbf{e}_l) - w(\mathbf{0})]}{\Delta^2},$$
(15)

where $\mathbf{0}$ is the zero tensor, and $\Delta$>0 is a small number. For each network constructed, we computed the strain energy density for 27 different deformations $\varepsilon$. These deformations were chosen to allow calculation of all 21 independent elastic moduli using equation (15) (e.g., assuming $w(\mathbf{0})$=0, calculating $C_{1111}$ requires finding $w$ for two different deformations, $\varepsilon$=$\Delta \mathbf{e}_1 \otimes \mathbf{e}_1$ and $\varepsilon$=$2\Delta \mathbf{e}_1 \otimes \mathbf{e}_1$). Thus, we determined the entire fully anisotropic elastic tensor for each network.

### 3.2 Numerical determination of $\rho_{ref}$, $\alpha$, and the five independent components of $K(n)$

Seventy networks with varying volume fractions were generated, and their elastic moduli numerically computed using the finite element method. Physiological volume fractions of filamentous actin within different types of cells and cytoplasmic regions are still under discussion. However, the average volume fraction within bovine aortic endothelial cells has been reported to be on the order of 1% [9]. Since we expect the volume fraction to be higher than the average within actin-rich regions such as the cortex or actin bundles, we constructed networks with volume fractions ranging from $\rho$=0-0.10. Note that "dangling" filament ends (i.e., filament segments attached to a cross-link at one end but to nothing at the other) were accounted for when quantifying volume fractions (these segments do not contribute to the elastic energy of the network). We determined $\rho_{ref}$ by finding the minimum volume fraction at which the elastic tensor was non-zero among all tested networks. We found $\alpha$ by plotting $C_{3333}$ as a function of volume fraction, and fitting this curve to a power law scaling of the form $(\rho-\rho_{ref})^{\alpha}$ (figure 5). Next, we computed the angular distribution $\alpha(\boldsymbol{n})$ for a single network ($\rho$=0.05), and determined the five independent components of $\boldsymbol{K}(\boldsymbol{n})$ using a direct search optimization to perform a least square fit, i.e., such that they minimized the norm of the difference between the network's elastic tensor calculated using finite elements, and the network's elastic tensor calculated using equation (2). Notably, performing this optimization for one network or many of them simultaneously did not appreciably change the value of $\boldsymbol{K}(\boldsymbol{n})$.

### 3.3 Predicting the elastic moduli of networks of varying volume fractions and degrees of anisotropy

Once $\rho_{ref}$, $\alpha$, and $\boldsymbol{K}(\boldsymbol{n})$ were determined, we sought to test the accuracy of equation (2) in predicting the elastic moduli of networks with different volume fractions and varying degrees of anisotropy. To this end, we generated additional networks for which the filament orientations were randomly sampled from a family of angular probability distributions $\omega_{\xi}$, $0 \leq \xi \leq 1$, constructed as a normalized linear combination of two real spherical harmonics $S_{00}$ and $S_{10}$, i.e.,

$$\omega_{\xi}(\boldsymbol{n}) = \frac{(1-\xi)S_{00}(\boldsymbol{n}) + \xi S_{10}(\boldsymbol{n})}{\int\limits_{S^2} [(1-\xi)S_{00}(\boldsymbol{n}) + \xi S_{10}(\boldsymbol{n})]dS} . \tag{16}$$

Here $S_{00}$ denotes the isotropic distribution function over the unit sphere, while $S_{10}$ stands for a dipole distribution aligned in the $\boldsymbol{e}_3$ direction (if $\phi$ is the angle between $\boldsymbol{e}_3$ and $\boldsymbol{n}$, then $S_{10}(\phi)=(\sqrt{3}\cos(\phi))/(2\sqrt{\pi})$). Constructing networks with $\xi$=0 resulted in networks with filaments isotropically aligned, while constructing networks with $\xi$=1 resulted in networks with filaments primarily aligned in the $\boldsymbol{e}_3$ direction. In general, constructing networks with $\xi$=1 resulted in networks with an approximately threefold increase in stiffness in $C_{3333}$ compared to $C_{1111}$ and $C_{2222}$. Note that the number of cross-links per network did not appreciably change with the degree of anisotropy tested here. We generated 240 networks with different volume fractions ($0 \leq \rho \leq 0.10$) and degrees of anisotropy ($0 \leq \xi \leq 1$), and computed their angular distributions $\alpha(\boldsymbol{n})$. Using equation (2), and the values of $\rho_{ref}$, $\alpha$, and $\boldsymbol{K}(\boldsymbol{n})$ found previously, predicted values for the elastic moduli were computed for each network. These values were then compared against the exact elastic moduli of each network computed with the finite element model. For each network the error between the predicted and exact values for the elastic moduli, $\boldsymbol{C}^{model}$ and $\boldsymbol{C}^{exact}$, respectively, was found as $\|\boldsymbol{C}^{model}-\boldsymbol{C}^{exact}\|_2/\|\boldsymbol{C}^{exact}\|_2$. Note that in general, we did not observe larger error in any one component of the elastic tensor for isotropic and anisotropic networks. Finally, in order to visualize the fourth-order elastic tensor $\boldsymbol{C}$, we generated polar plots of the quantity $C_{ijkl}n_in_jn_kn_l$ as a function of direction $\boldsymbol{n}$. Physically, $C_{ijkl}n_in_jn_kn_l$ gives the normal component of the traction in direction $\boldsymbol{n}$ generated by a uniaxial unit strain in the same direction.

## 4 Results

The numerically determined values of $\rho_{ref}$ and $\alpha$ were 0.01 and 1.6, respectively. The five independent components of $\boldsymbol{K}(\boldsymbol{n})$, in the form of equation (8), were $K_{1111}(\boldsymbol{e}_3)$=5.4MPa, $K_{1122}(\boldsymbol{e}_3)$=9.7MPa, $K_{1133}(\boldsymbol{e}_3)$=13.0MPa, $K_{3333}(\boldsymbol{e}_3)$=3.8GPa and $K_{1313}(\boldsymbol{e}_3)$=37.0MPa (figure 6). Using these values, equation (2) predicted the elastic tensors of networks generated with constant volume fraction ($\rho$=0.05) and varying degrees of anisotropy ($0 \leq \xi \leq 1$) to within 14.5±0.6% (mean±SE, n=100 networks[1]). As figures 7 and 8 show, there seems to be only a minimal dependence of the error on anisotropy. Next, we used equation (2) to compute predicted values for the elastic moduli of both isotropic ($\xi$=0) and anisotropic ($\xi$=1) networks generated over a range of different volume fractions ($0 \leq \rho \leq 0.10$). The results are shown in figure 9. For both isotropic and anisotropic networks, the error was found to be relatively constant with respect to the volume fraction, except for volume fraction values approaching $\rho_{ref}$, for which the errors increased rapidly. For example, for networks with volume fractions far from $\rho_{ref}$ ($0.03 < \rho \leq 0.10$), the error was relatively small ($\xi$=0: 12.6±0.9%, n=48; $\xi$=1: 12.1±0.9%, n=48), while for networks with volume fractions greater than but near $\rho_{ref}$ ($\rho = \rho_{ref}$-0.03), the error was substantially larger ($\xi$=0: 52.8±5.8%, n=13; $\xi$=1: 57.1±6.7%, n=14). For networks with volume fractions smaller than $\rho_{ref}$ the elastic tensor was identically zero, reflecting the fact that $\rho_{ref}$ is a good estimate of the percolation threshold for these networks.

## 5 Discussion

We proposed here a novel class of models for the homogenized linear elastic response of cross-linked polymer networks. The model requires determination of only seven independent parameters: five constants to construct a fourth-order tensor valued function $\boldsymbol{K}(\boldsymbol{n})$ that incorporates different aspects governing network mechanical behavior (e.g., filament elasticity, cross-link type, etc.), and two constants, $\rho_{ref}$ and $\alpha$, to account for nonlinear scaling of network elastic moduli with filament density.

We demonstrated the applicability and validity of these models by numerically determining $\rho_{ref}$, $\alpha$, and the five independent constants used to construct $\boldsymbol{K}(\boldsymbol{n})$ for a particular class of network. We obtained a value of 1.6 for $\alpha$, indicating that the elastic moduli in our three-dimensional networks scaled with filament density as $\sim(\rho-\rho_{ref})^{1.6}$. This power law dependence is similar to the scaling exponent of 1.8 found in two-dimensional networks of rods with stiff cross-links near the percolation threshold [15]. Studies of three-dimensional systems of rods [47, 48, 49, 50] suggest that the critical volume fraction necessary for percolation for an infinite system of randomly oriented, straight slender rods takes the form [44, 51, 52] $d/(2l) \sim 8/(700) \sim 0.01$, same as the value of 0.01 we obtained for $\rho_{ref}$. Out of the five different values that determine $\boldsymbol{K}(\boldsymbol{n})$, we found that $K_{3333}(\boldsymbol{e}_3)$ was two or three orders of magnitude greater than the other components. This indicates that networks in the class considered herein predominantly absorb deformation by stretching of the filaments, instead of from bending or torsion. The nonzero value of the other coefficients indicates that either some degree of nonaffinity exists in the deformed position of the cross-links, and/or reflects the presence of cross-links that do not freely rotate.

---

[1] SE: standard error, n: number of samples

When we calculated the eigenvalues of $K(e_3)$ over all symmetric second-order tensors, we found them to be all positive. This implies that the homogenized linear elastic moduli will be positive definite over all symmetric second-order tensors for any volume fraction $\rho > \rho_{ref}$ and angular probability density $\omega(n)$, a fact that renders the deformations of any of the resulting homogenized models elastically stable.

Once we numerically determined $K(n)$, $\rho_{ref}$ and $\alpha$, we used equation (2) to compute the predicted values for the elastic tensors of networks over a range of volume fractions ($0 \leq \rho \leq 0.10$), and with varying degrees of anisotropy ($0 \leq \xi \leq 1$). We discuss here several trends in the error in the predicted homogenized elastic moduli with respect to volume fraction and anisotropy. First, although the parameters in equation (2) were numerically derived from isotropic networks, the model generally predicted the elastic moduli of isotropic and anisotropic networks equally well. However, we expect that the error may be substantial for networks with higher degrees of anisotropy than those tested in this study. For example, for a theoretical network in which every filament is aligned in the $e_3$ direction, the model predicts finite axial stiffness in the transverse directions, which is obviously not correct. Second, we found that the error was large for networks with volume fractions near $\rho_{ref}$, regardless of the degree of anisotropy. This is indicative of the fact the elastic moduli near $\rho_{ref}$ may scale differently with volume fraction at these very low values. Finally, we found that the error was relatively constant for volume fractions far from $\rho_{ref}$. It would not be unexpected for the error to increase substantially for networks with volume fractions larger than those tested in this study. At the volume fractions studied here the mechanical response of the network is determined mainly by the stretching mechanical behavior of the filaments. However, at higher volume fractions other phenomena may become important, as some two-dimensional studies indicate [10, 13]. Alternatively, simple inspection of the previous example consisting in a cubic microstructure with filaments that only have bending stiffness suggests that at higher volume fractions the bending response may become dominant, since in this case the stiffness scales quadratically with the volume fraction. We have not yet been able to experimentally verify this regime due to the large computational cost associated with high density networks in three dimensions.

Overall, once calibrated the model predicted the elastic moduli for these networks over a range of volume fractions and degrees of anisotropy. In fact, the numerical results are surprisingly good, since the variation of the moduli with the anisotropy parameter was well captured, despite the fact that the model parameters were determined with isotropic networks. More generally, we expect the proposed model, with perhaps some minor extensions, to be useful in different regimes and for other types of networks not considered herein as well, if calibrated accordingly. This, of course, has to be explicitly studied for each network type, e.g., different cross-linker microscopic properties, prestress in the network and length distribution of actin filaments, among others.

The distinguishing feature of this class of models is that they can account for filament angular distribution and volume fraction, two microstructural parameters that can be estimated from light microscopy data. For example, although individual filaments cannot be resolved in fluorescent images (a typical pixel/voxel size in fluorescent microscopy is 0.2~0.4μm, whereas actin filaments are approximately 8nm in diameter), one can approximate the number and orientations of filaments within each image voxel. Specifically, the volume fraction of filaments within each voxel can be estimated from the voxel intensity, while their orientations can be estimated from the local image texture [53, 54] to give an approximate angular distribution of filaments [53]. Thus, the approach presented here may be used to construct cell structural models that account for the anisotropic and heterogeneous distribution of actin filaments, something that has yet to be explored and validated against appropriate experiments. Despite the facts that the actin cytoskeleton is a highly dynamic structure capable of remodeling, and that cells can undergo large deformations, the linearized mechanical response studied herein may provide important insight for the interpretation of a wide class of microrheological experiments.

## 6 Acknowledgments

We would like to acknowledge grant AR45989 from the National Institutes of Health, the National Science Foundation Graduate Fellowship, and the Veterans Affairs Palo Alto Bone and Joint Center for funding.

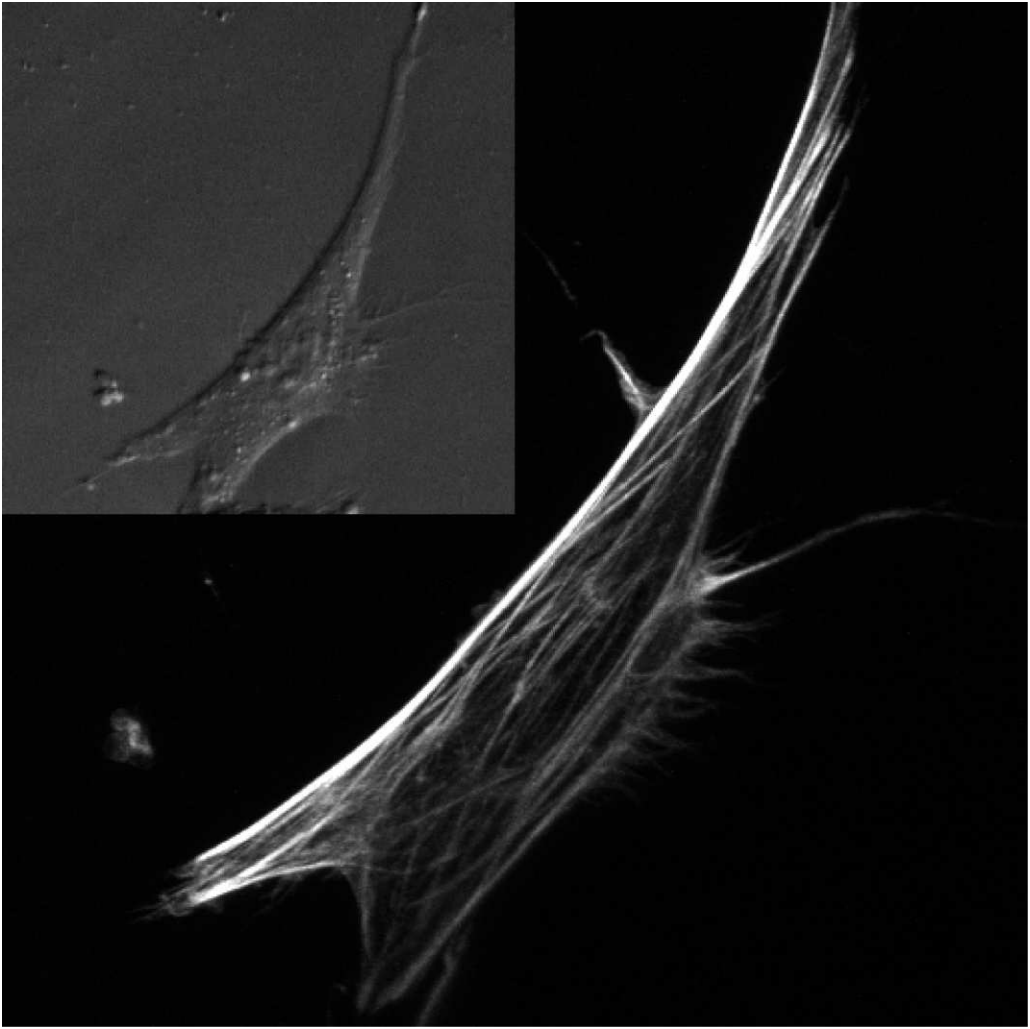

**Figure 1.** Fluorescent image of the actin cytoskeleton of a MC3T3-E1 osteoblastic (bone) cell. Microstructurally, the actin cytoskeleton is highly heterogeneous in both the number and orientation of filaments at each point. The inset contains a brightfield image showing the same cell.

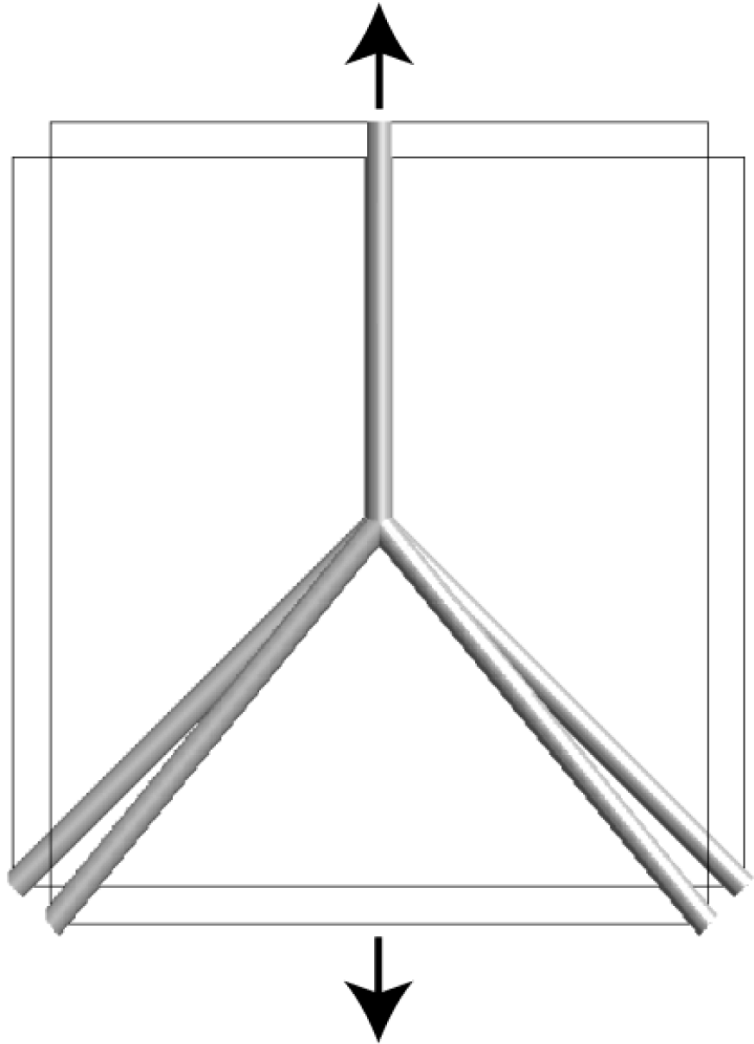

**Figure 2.** Schematic depicting a simple, non-periodic network in which pulling the vertical filament in the axial direction results in a contraction in the transverse direction. In this case, the filament transmits forces along its own direction and, through its connections to the rest of the network, in transversal directions as well.

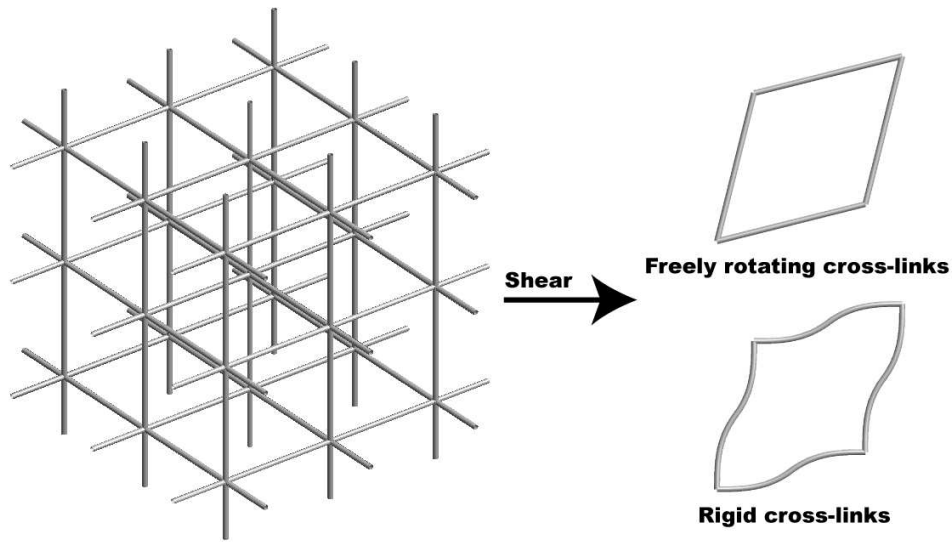

**Figure 3.** Three-dimensional network with prismatic microstructure (left) and representative unit cells when undergoing shear (right). If the cross-links are rigid, and the network undergoes shear, the filaments in the plane of shear undergo bending. In contrast, if the cross-links freely rotate, the filaments only undergo axial stretch.

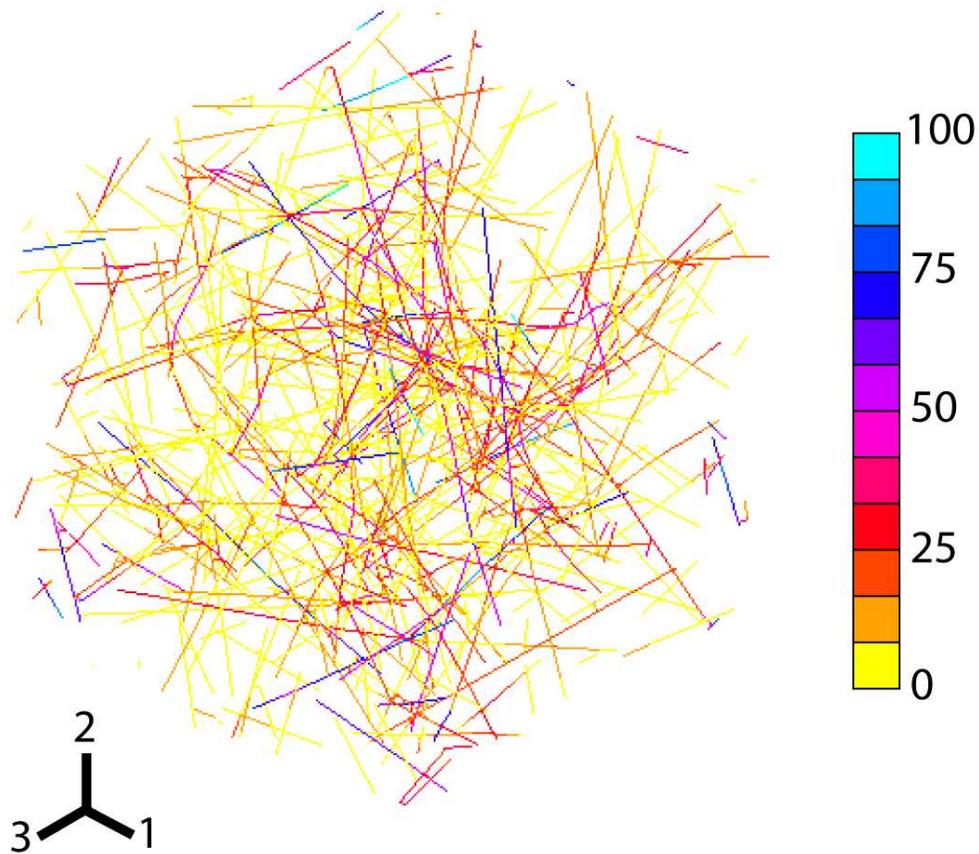

**Figure 4.** Typical three dimensional, cross-linked actin network used in this study. The actin filaments and cross-links are modeled as elastic rods. Shown is the Von Mises stress distribution calculated using the finite element method when the network undergoes shear in the 1-2 plane. Units are in MPa.

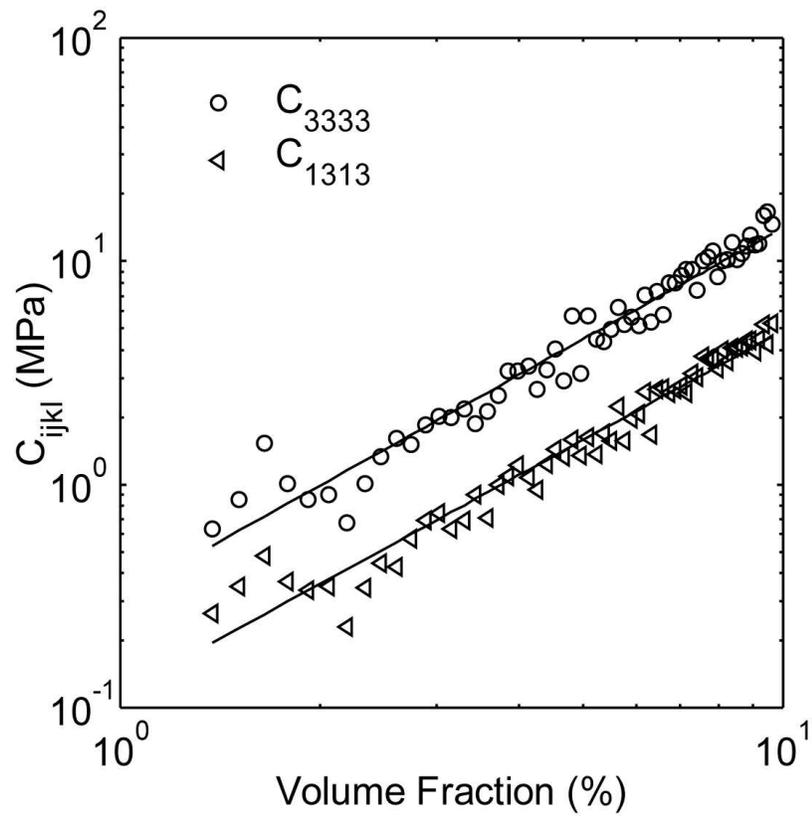

**Figure 5.** Log-log plot of elastic moduli $C_{3333}$ and $C_{1313}$ versus volume fraction $\rho$. The elastic moduli scale as $\sim\rho^{1.6}$ (black lines). Thus, a value of $\alpha$=1.6 was used in equation (2).

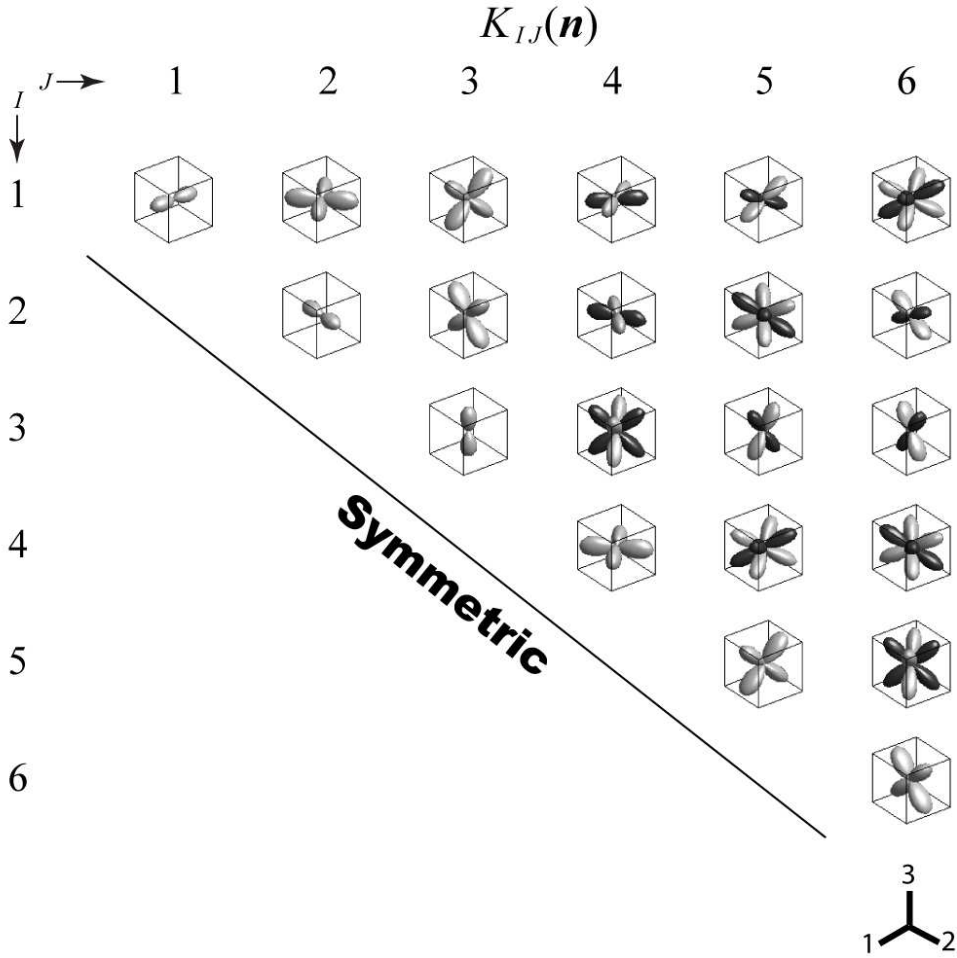

**Figure 6.** Polar plots of the components of $K_{IJ}(\boldsymbol{n})$, the 6x6 matrix representation of $\boldsymbol{K}$. The plot in the $I$th row and $J$th column is the polar plot of $K_{IJ}(\boldsymbol{n})$. Positive values are represented in gray, negative values are represented in black. The magnitude of $K_{IJ}$ in the direction $\boldsymbol{n}$ indicates the degree to which filaments oriented in the direction $\boldsymbol{n}$ contribute to elastic modulus $C_{IJ}$. For example, $K_{11}$ is largest in the $\boldsymbol{e}_1$ direction, indicating that filaments oriented in this direction contribute the most to $C_{11}$. Note that for clarity, the components were not all plotted on the same scale.

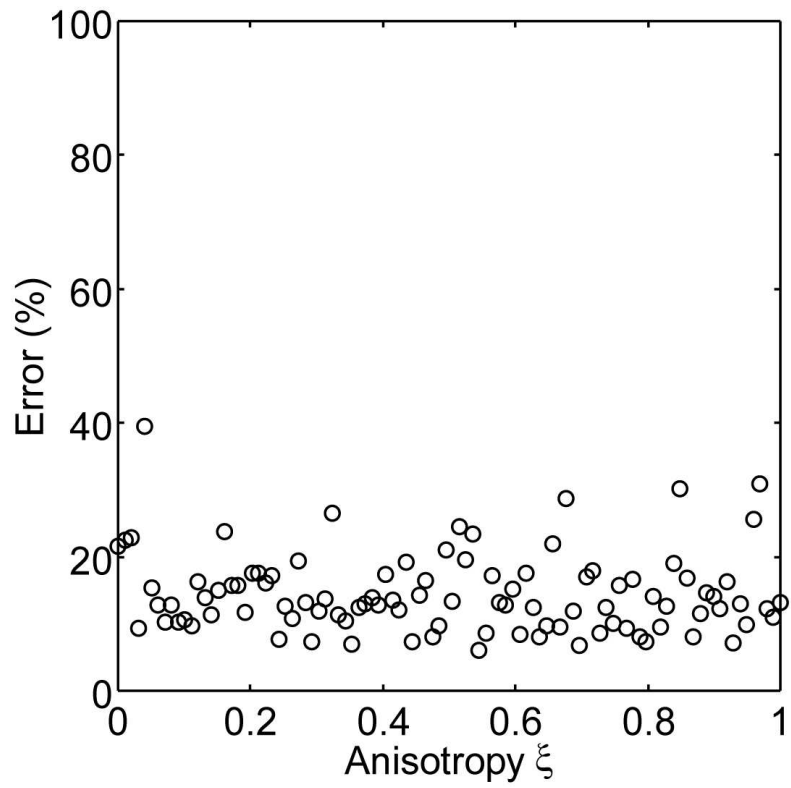

**Figure 7.** Error in the predicted elastic moduli as a function of the degree of anisotropy $\xi$. A minimal dependence of the error on anisotropy is observed.

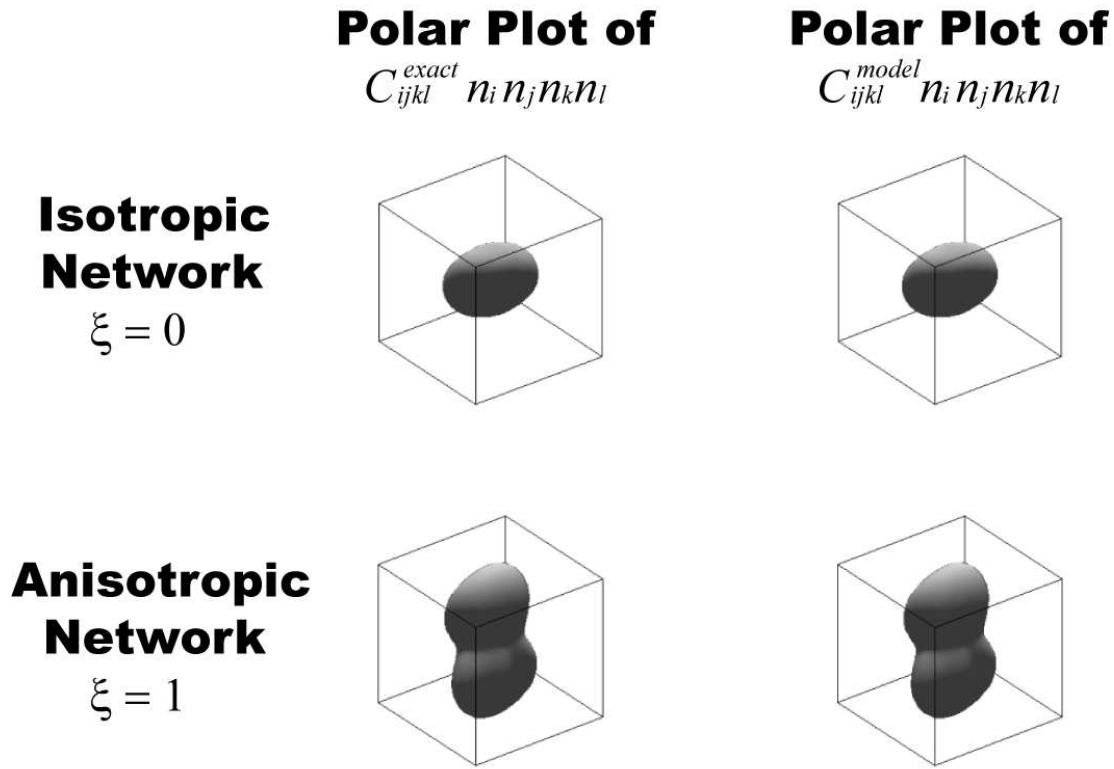

**Figure 8.** Polar plots of $C_{ijkl}n_in_jn_kn_l$ as a function of direction $\boldsymbol{n}$ using elastic moduli predicted by the model ($\boldsymbol{C}^{model}$, right column), and generated using elastic moduli calculated using the finite element method ($\boldsymbol{C}^{exact}$, left column). The top row represents an isotropic network ($\xi$=0), while the bottom row represent an anisotropic network ($\xi$=1). The plots in the left and right columns appear identical, indicating the elastic moduli calculated using the finite element method are well matched by those calculated using the model, for both isotropic and anisotropic networks. The plot boundaries span 30MPa.

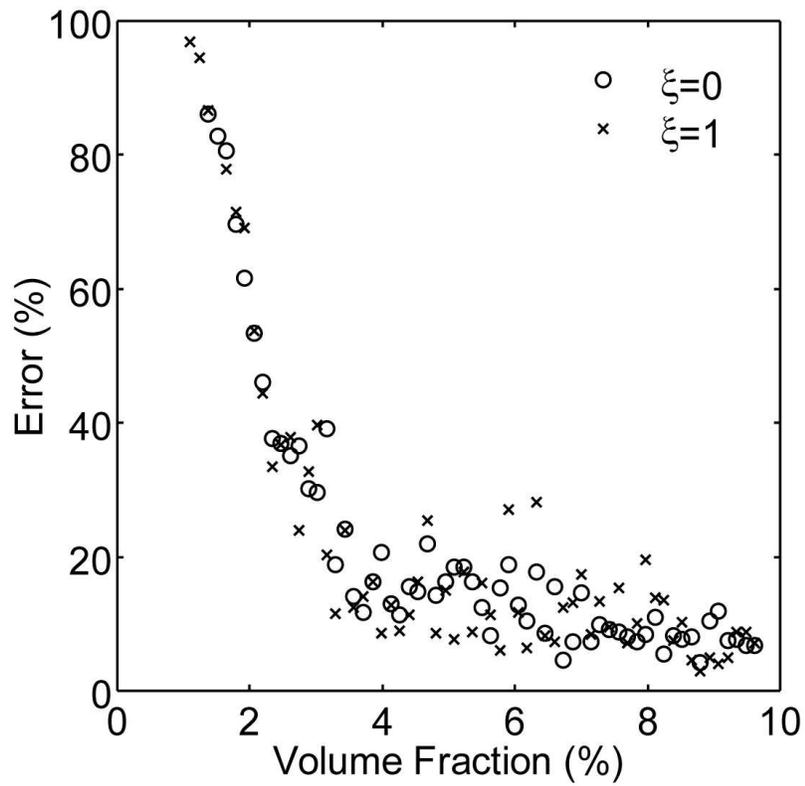

**Figure 9.** Error in the predicted elastic moduli as a function of volume fraction $\rho$ for networks with different degrees of anisotropy ($\xi$=0,1). For both isotropic and anisotropic networks, the error was constant far from $\rho_{ref}$~0.01, but increased as the volume fraction approached $\rho_{ref}$.